\documentclass[a4paper,twoside,12pt]{JHEP3}
\usepackage{epsfig} 
\usepackage{amssymb} 
\usepackage{amsmath} 
 
\newcommand{\identity}{{\rlap{1} \hskip 1.6pt \hbox{1}}} 
 
\def\Tr{{\rm Tr}}

\preprint{FIRENZE DFF-408/10/03\\ 
LYCEN-2003-39}

\title{Little Higgs Models and Precision Electroweak Data} 
 
\author{R. Casalbuoni\\Dipartimento di 
 Fisica, Universit\`a di Firenze \\ 
and I.N.F.N., Sezione di Firenze,  
I-50019 Firenze, Italy\\ E-mail: \email{casalbuoni@fi.infn.it}} 
\author{A. Deandrea\\Universit\'e de Lyon 1, Institut de Physique  
Nucl\'eaire,\\ 
4 rue E.~Fermi, F-69622 Villeurbanne Cedex, France\\  E-mail:  
\email{deandrea@ipnl.in2p3.fr}} 
\author{M. Oertel\\CEA DPTA-SPN, B.P. 12,  
F-91680 Bruy\`eres-le-Ch\^atel, France\\ E-mail:  
\email{oertel@cea.fr}} 
 
\abstract{We study the low energy limit of a Little Higgs model with 
custodial symmetry.  The method consists in eliminating the heavy 
fields using their classical equations of motion in the infinite mass 
limit. After the elimination of the heavy degrees of freedom we can 
directly read off deviations from the precision electroweak data. We 
also examine the effects on the low energy precision experiments.} 
 
 
\begin{document} 
 
\section{Introduction} 
 
One of the major problems affecting the Standard Model (SM) is the so 
called hierarchy problem, that is the enormous difference between the 
electroweak and the Planck scales. In fact, since within the SM the 
Higgs gets a quadratically divergent contribution to its mass, that 
would imply a Higgs mass of the order of the Planck mass. On the other 
hand LEP and SLC, together with other low energy experiments, have 
clearly shown that the physics behind the SM is perturbative in 
nature. This means that the Higgs mass cannot be very large. This 
requires fine tuning from the Planck scale to the electroweak 
scale. Clearly the situation is not satisfactory and there have been 
various proposals to avoid the problem. One is supersymmetry, where 
the quadratic divergence in the Higgs mass is cancelled by the 
fermions. Another one is technicolor, where the problem is solved 
lowering the relevant scale from the Planck mass to values of the 
order of TeV. Recently it has been proposed~\cite{Arkani-Hamed:2001nc} 
to consider the Higgs fields as Nambu Goldstone Bosons (NGB) 
\cite{Dimopoulos:1981xc} of a global symmetry which is spontaneously 
broken at some higher scale $f$ by an expectation value. The Higgs 
field gets a mass through symmetry breaking at the electroweak 
scale. However since it is protected by the approximate global 
symmetry it remains light. An important point is that the cancellation 
of the quadratic divergence is realized between particles of the same 
statistics. 
 
Of course all models containing new physics are highly constrained by the 
electroweak precision tests. Aim of this paper is to consider the 
electroweak precision data constraints on Little Higgs models by 
using a general method based on the effective Lagrangian approach. 
The idea is simple: we eliminate the heavy fields from the 
Lagrangian via their classical equations of motion in the limit of 
infinite mass, which means in practice that their mass must be 
much bigger than $m_W$. We obtain an effective Lagrangian in terms 
of the Standard model fields, from which we can directly read off the 
deviations. 
 
We shall consider in detail in the following a model which exhibits an 
approximate $SU(2)$ custodial symmetry. 
The method is quite general and can be easily applied to other 
models. Similar ideas are discussed in \cite{othlh} for the littlest 
Higgs model and a class of other models. We study the electroweak  
precision constraints in terms of the $\epsilon$'s parameterization  
\cite{alt}. In order to fix our notations we shall briefly review the  
littlest Higgs model in section~\ref{epsilons}.  
Section~\ref{epscust} will be devoted to the study of electroweak 
corrections within a model which has an approximate custodial 
symmetry and in section~\ref{lowenergy} we will investigate the 
low energy precision data within both models. In Appendix A we give the  
expressions for the couplings necessary for the evaluation of $g-2$ and  
in Appendix B those for the weak charge. 
 
\section{Littlest Higgs model} 
\label{epsilons}  
In order to fix our notations and make contact with the existing literature 
we briefly consider here the littlest Higgs model.  
Note that our analysis in terms of the $\epsilon$-parameters focuses 
only on the oblique corrections. More complete investigations of the 
constraints imposed by electroweak precision data on the littlest 
Higgs model have already been presented in the literature, using 
various methods~\cite{othlh}. We discuss it here mainly in 
order to show the differences with the model incorporating custodial 
symmetry in the following section.  
 
The model is based on a $SU(5)$ symmetry with a $[SU(2)\times U(1)]^2$ 
subgroup gauged.  This symmetry is broken down to $SO(5)$ by a vev of 
the order $f$.  
This vev also breaks the gauge symmetry to $SU(2)_W\times U(1)_Y$. This 
symmetry breaking patterns leads to 14 Goldstone bosons. Four of 
them are eaten up by the gauge bosons of the broken gauge group.  
The Goldstone boson matrix contains a Higgs doublet and a triplet under the unbroken SM gauge group.  
%
More details about this specific model and the corresponding notations 
can be found in Ref.~\cite{othlh,Han:2003wu}. 
 
The kinetic term for the scalar sigma model fields $\Sigma$ is given by 
\begin{equation} 
\mathcal{L}_\mathit{kin} = \frac{1}{2} \frac{f^2}{4} \Tr[D_\mu \Sigma D^\mu \Sigma]~, 
\end{equation} 
with the covariant derivative defined as 
\begin{equation} 
D_\mu \Sigma = \partial_\mu \Sigma - i (A_\mu \Sigma + \Sigma A_\mu^T)~. 
\end{equation} 
With $A_\mu$ we denote the gauge boson matrix: 
\begin{equation} 
A_\mu = g_1 W_\mu^{1a} Q^{a}_1 + g_2 W_\mu^{2a} Q^{a}_2 + g'_1 B_\mu^1 Y_1 + g'_2 B_\mu^2 Y_2~, 
\end{equation} 
where the $Q^a_i$ are the generators of the two $SU(2)$ groups and the 
$Y_i$ are the generators of the two $U(1)$ groups, respectively. After 
symmetry breaking the gauge boson matrix can be diagonalized by the 
following transformations: 
\begin{alignat}{4} 
W&= s W_1 + c W_2 \qquad & W' &= -c W_1 + s W_2 \nonumber \\ 
B&= s' B_1 + c' B_2 \qquad & B' &= -c' B_1 + s' B_2~. 
\end{alignat} 
$s, c, s',$ and  $c'$ denote the sines and cosines of two mixing angles, 
respectively. They can be expressed with the help of the coupling 
constants: 
\begin{eqnarray} 
\nonumber 
c' &=& g'/g'_2 \hspace{0.5in} s' = g'/g'_1 \\ 
c &=& g/g_2 \hspace{0.5in} s = g/g_1~, 
\end{eqnarray} 
with the usual SM couplings $g,g'$, related to $g_1$, $g_2$, 
$g_1'$ and $g_2'$ by 
\begin{equation} 
\frac 1 {g^2}=\frac 1{g_1^2}+\frac 1{g_2^2},~~~~~~~ \frac 1 
{{g'}^2}=\frac 1{{g_1'}^2}+\frac 1{{g_2'}^2}~.\label{eq:10} 
\end{equation} 
 
The equations of motion for the heavy gauge bosons can now easily be 
obtained from the complete Lagrangian. 
We neglect, at the lowest order in the momenta,  
derivative contributions, i.e., the contributions from the kinetic 
energy vanish. Up to the order $v^2/f^2$ we obtain: 
\begin{eqnarray} 
W'^{\pm\mu} &=& \frac{cs}{2} (c^2 - s^2) \frac{v^2}{f^2}\, W^{\pm\mu} 
- \frac{4 c^3 s}{\sqrt{2}g f^2}\left( 
J^{\pm\mu} - 
(1-c_L) J^{\pm\mu}_3\right) \\ 
W'^{3\mu} &=& \frac{cs}{2} (c^2 - s^2) \frac{v^2}{f^2} (W^{3\mu} + \frac{g'}{g} \,B^\mu) - \frac{4 c^3 s}{g f^2} \left( 
J^{0\mu} -  s_L^2 \bar{t}_L \gamma^\mu t_L\right)\\ 
B'^{\mu} &=&  2c's' (c'^2 - s'^2) \frac{v^2}{f^2}(\frac{g}{g'} \, W^{3\mu} + B^\mu) \nonumber \\ && +   
 \frac{4 c' s' }{g' f^2} \left[(3 c'^2 - 2 s'^2) (J^\mu_\mathit{em} + J^{0\mu}) - \frac{5}{2} c'^2 
s_L^2 \bar{t}_L \gamma^\mu t_L  
-s_R^2 \bar{t}_R \gamma^\mu t_R \right]~, 
\end{eqnarray} 
where we have used the notation of Ref.~\cite{Han:2003wu} for the 
diagonalisation of the top sector.  
The currents are defined as usual and $J^{\pm\mu}_3$ describes the 
current of quarks of the third generation, i.e., bottom and top. 
Due to the mixing with the heavy top for the quarks of the third generation the neutral current is modified, too: 
\begin{eqnarray} 
-Z_\mu \sqrt{g^2 + g'^2} \Bigg[ J^{0\mu}(1-\frac{v^2}{f^2} (\frac{c^2}{2} (c^2-s^2) + \frac{2}{5} ( 3 c'^2-2 s''^2)( c'^2-s'^2)) )&& 
\nonumber \\+  J^\mu (\frac{g'^2}{g^2 + g'^2} - \frac{2 g}{5 g'} \frac{v^2}{f^2} (3 c'^2-2 s'^2) (c'^2-s'^2))&& \nonumber \\ 
- \frac{s_L^2}{2} \bar{t}_L \gamma^\mu t_L (1-\frac{v^2}{f^2} (\frac{c^2}{2} (c^2-s^2) + 2 (c'^2-s'^2) )) 
+ \frac{v^2}{f^2} \frac{s_R^2}{5} (c'^2-s'^2) \bar{t}_R \gamma^\mu t_R\Bigg]&& \; . 
\end{eqnarray} 
However this modification is irrelevant for the analysis of the 
precision electroweak data, as $t{\bar t}$ production at LEP was 
kinematically not possible. Therefore we we will discard the $t{\bar 
t}$ correction in the following evaluations. 
The heavy Higgs particles completely decouple at that order.  
 
To determine the $\epsilon$-parameters we proceed in the same way as 
in Ref.~\cite{Anichini:1994xx} and first look at the modification 
to $G_F$. We have two types of modifications: one directly from 
the mixing of the heavy $W'$ bosons to the coupling of the charged 
current and the second one form the contribution of the charged 
current to the equations of motion of the heavy gauge bosons. The 
input parameters in the analysis of the electroweak data are the 
Fermi constant $G_F$, the mass of the $Z$ vector boson $m_Z$ and 
the fine--structure coupling $\alpha(m_Z)$. In terms of the model 
parameters we obtain: 
\begin{equation} 
\frac{G_F}{\sqrt{2}} = 
\frac{\alpha \pi (g^2 + g'^2)}{2 
g^2 g'^2 m_Z^2} \left(1- c^2 
(c^2-s^2)\frac{v^2}{f^2} + 2 c^4 \frac{v^2}{f^2}- \frac{5}{4} 
(c'^2-s'^2)^2 \frac{v^2}{f^2}\right)\; . 
\end{equation} 
We define the Weinberg angle  as~\cite{Anichini:1994xx}: 
\begin{equation} 
\frac{G_F}{\sqrt{2}} = \frac{\alpha \pi}{2 s_\theta^2 c_\theta^2 m_Z^2}~. 
\label{weinberg} 
\end{equation} 
In terms of the model parameters the mass of the $Z$-boson is given by 
\begin{equation} 
m_Z^2 = (g^2 + g'^2) \frac{v^2}{4} \left[ 1-\frac{v^2}{f^2} 
\left( \frac{1}{6} + \frac{(c^2-s^2)^2}{4} 
+ \frac{5}{4} (c'^2-s'^2)\right) + 8 \frac{v'^2}{v^2}\right]~, 
\end{equation} 
whereas the $W$-mass is 
\begin{equation} 
m_W^2 = \frac{g^2 v^2}{4} \left[ 1- \frac{v^2}{f^2} 
    \left(\frac{1}{6} +\frac{(c^2-s^2)^2}{4}\right) 
       + 4 \frac{v'^2}{v^2}\right]~. 
\end{equation} 
The expression for 
the $Z$-mass can be used to determine the value of $v$ for a given 
ratio $v/f$.  
 
Our result for the corrections to the $\epsilon_i$ parameters to the 
order $v^2/f^2$ is given by: 
\begin{eqnarray} 
\epsilon_1 &=& - \frac{v^2}{f^2}\left( \frac{5}{4}(c'^2-s'^2)^2  
+ \frac{4}{5} (c'^2-s'^2) (3 c'^2-2 s'^2) + 2 c^4 \right) 
+ 4 \frac{v'^2}{v^2} \\ 
\epsilon_2 &=& - 2 c^4 \frac{v^2}{f^2}\\ 
\epsilon_3 &=& - \frac{v^2}{f^2}\left(\frac{1}{2 } c^2 (c^2-s^2)  
+ \frac{2}{5} (c'^2-s'^2) (3 c'^2 - 2s'^2)\frac{c_\theta^2}{s_\theta^2}\right) 
\end{eqnarray} 
Notice that the corrections, as they should, depend only on the 
parameters $c$, $c'$, $v/f$ and $v'/v$.  Using the values of the 
$\epsilon_i$ parameters given in \cite{Altarelli:2001wx} 
\begin{equation} 
\begin{array}{c} 
\epsilon_1=~~(5.1\pm  1.0)\times 10^{-3}\\ 
\epsilon_2=(-9.0\pm 1.2)\times 10^{-3}\\ 
\epsilon_3=~~(4.2\pm  1.0)\times 10^{-3} 
\end{array} 
\end{equation} 
one can easily compare the model with data. 
These values only assume lepton universality and the derivation of the 
$\epsilon_i$ is otherwise completely model independent. 
 
There are no stringent limits on the values of the triplet vev, such 
that a priori $v'/v$ can be treated as completely arbitrary. The 
authors of Ref.~\cite{Han:2003wu} obtain a bound of \mbox{$v'^2/v^2 < 
v^2/(16 f^2)$} in order to maintain a positive definite triplet mass 
for the Higgs. Throughout our analysis we assumed as a reasonable 
guideline $v'/v$ being at least of the order $v/f$. 
 
Our results are in agreement with those reported in the literature 
\cite{othlh}. In particular for large values of $v/f$ the allowed 
regions are very small, whereas for small values practically the 
entire parameter space is excluded.  For large values of $v/f$ this is 
mainly due to the fact that this model exhibits no custodial symmetry 
and that it is therefore difficult to satisfy the experimental 
constraint on $\epsilon_1$ without fine tuning of the parameters. For 
small values of $v/f$ we approach the SM limit which itself is not in 
agreement with the values for the $\epsilon$-parameters. A variation 
of the triplet vev does in this case only slightly modify the results 
but does not change the general conclusions. 
 
\section{Little Higgs with custodial $SU(2)$} 
\label{epscust} We now examine a ``little Higgs'' model which 
has an approximate custodial $SU(2)$ symmetry \cite{Chang:2003un}. 
The model is based on a $SO(9)/[SO(5)\times SO(4)]$ coset space, 
with $SU(2)_L\times SU(2)_R \times SU(2) \times U(1)$ subgroup of 
$SO(9)$ gauged.  
 
One starts with an orthogonal symmetric nine by nine matrix, representing a 
nonlinear sigma model field $\Sigma$ 
which transforms under an $SO(9)$ rotation 
by $\Sigma \to V\Sigma V^T$.  To break the $SO(4)$'s to the diagonal, one can 
take $\Sigma$'s vev to be 
\begin{eqnarray} 
\langle \Sigma \rangle=\left(\begin{array}{ccc} 
0 & 0 & \identity_4 \\ 
0 & 1 & 0 \\ 
\identity_4 & 0 & 0 \end{array}\right) 
\end{eqnarray} 
breaking the $SO(9)$ global symmetry down to an $SO(5)\times SO(4)$ subgroup. 
This coset space has $20 = (36-10-6)$ light scalars. Of these 
20 scalars, 6 will be eaten in the higgsing of the gauge groups down to 
$SU(2)_W \times U(1)_Y$. The remaining 14 scalars are : a single higgs 
doublet $h$, an electroweak singlet $\phi^0$, and three triplets $\phi^{ab}$ 
which transform under the $SU(2)_L \times SU(2)_R$ diagonal 
symmetry as 
\begin{eqnarray} 
h: (\mathbf{2}_L,\mathbf{2}_R) \hspace{0.5in} 
\phi^0 : (\mathbf{1}_L,\mathbf{1}_R) 
\hspace{0.5in} \phi^{ab} : (\mathbf{3}_L,\mathbf{3}_R). 
\end{eqnarray} 
These fields can be written 
\begin{eqnarray} 
\Sigma = e^{i\Pi/f}\langle\Sigma\rangle e^{i\Pi^T/f} = 
e^{2i\Pi/f}\langle\Sigma\rangle 
\end{eqnarray} 
with 
\begin{eqnarray} 
\Pi = \frac{-i}{4}\left(\begin{array}{ccc} 
0_{4\times 4} & \sqrt{2} \vec{h} & -\Phi \\ 
-\sqrt{2} \vec{h}^T & 0_{1\times 1} & \sqrt{2} \vec{h}^T \\ 
\Phi & -\sqrt{2} \vec{h} & 0_{4\times 4} 
\end{array}\right) 
\end{eqnarray} 
where the Higgs doublet $\vec{h}$ is written as an $SO(4)$ vector; 
the singlet and triplets are in the symmetric four by four 
matrix $\Phi$ 
\begin{eqnarray} 
\Phi = \phi^0 + 4\phi^{ab}\: T^{l\,a} T^{r\,b}\, , 
\end{eqnarray} 
and the would-be Goldstone bosons that are eaten in the higgsing to 
$SU(2)_W\times U(1)_Y$ are set to zero in $\Pi$.  The global 
symmetries protect the higgs doublet from one-loop quadratic divergent 
contributions to its mass.  However, the singlet and triplets are not 
protected, and are therefore heavy, in the region of the TeV 
scale. The theory contains the minimal top sector with two extra 
coloured quark doublets and their charge conjugates.  Further details 
and formulas can be found in \cite{Chang:2003un}. 
 
The kinetic energy for the pseudo-Goldstone bosons is 
\begin{eqnarray} 
{\cal L}_{kin} = \frac{f^2}{4}\Tr\left[D_\mu\Sigma D^\mu\Sigma\right] 
\end{eqnarray} 
and the covariant derivative is 
\begin{eqnarray} 
D_\mu \Sigma = \partial_\mu\Sigma +i\left[A_\mu,\Sigma \right] 
\end{eqnarray} 
where the gauge boson matrix $A_\mu$ is defined as 
\begin{eqnarray} 
A \equiv g_L W^{la}_{SO(4)} \tau^{l\,a} + g_R W^{ra}_{SO(4)} \tau^{r\,a} + g_2  W^{la} \eta^{l\,a}+ g_1 W^{r3} \eta^{r\,3}. 
\end{eqnarray} 
The $\tau^a$ and $\eta^a$ are the generators of two SO(4) 
subgroups of SO(9). For details see Ref.~\cite{Chang:2003un}. 
 
The vector bosons can be diagonalized with the following 
transformations: 
\begin{eqnarray} 
B &=&c' W^{r3} - s' W_{SO(4)}^{r3} \hspace{0.5in} 
B'= W'{}^{\,r3} = s' W^{r3} + c'  W_{SO(4)}^{r3}\\ 
W^a &=& c W^{la} +  s W_{SO(4)}^{la} \hspace{0.5in} 
W'{}^a=W'{}^{\,la} = -s W^{la} + c W_{SO(4)}^{la} 
\end{eqnarray} 
where the cosines and the sines of the mixing angles can be written in 
terms of the couplings 
\begin{eqnarray} 
\nonumber 
c' &=& g'/g_1 \hspace{0.5in} s' = g'/g_R \\ 
c &=& g/g_2 \hspace{0.5in} s = g/g_L. 
\end{eqnarray} 
Again $g$ and $g'$ are defined in terms of $g_1$, $g_R$ and $g_2$, $g_L$ respectively, as in equation (\ref{eq:10}). 
Neglecting possible corrections to the currents from the top sector, 
we obtain the following equations of motion up to the order $v^2/f^2$ 
\begin{eqnarray} 
W^{\prime \, 1,2} &=& -\frac{v^2\, c s}{2 f^2}\left( c^2-s^2 \right)\, W^{1,2} 
+\frac{s^3 c}{f^2 \, g}\, J^{1,2}\\ 
W^{\prime \, 3} &=& -\frac{v^2\, c s}{2f^2}\left( c^2-s^2 \right)\, ( W^3 - \frac{g'}{g}\, B) +\frac{s^3 c}{f^2 \, g}\, J^3\\ 
B^{\prime}&=&\frac{v^2\, c' s'}{2f^2}\left( c^{\prime\, 2}-s^{\prime\, 2} \right)(\frac{g}{g'}\, W^3 - B) +\frac{s^{\prime \, 3} c'}{f^2\, g'}\, J^0\\ 
W_R^{1,2} &=& \frac{v^2}{2f^2}\, W^{1,2}~. 
\end{eqnarray} 
We now proceed in exactly the same way as in the previous section and 
look first at the modifications to $G_F$.  The expression for $G_F$ in 
terms of the model parameters is %
\begin{equation} 
\frac {G_F}{\sqrt{2}} = \frac{\alpha \pi (g^2 + g'^2)^2}{2 g^2 
g'^2}\left(1 + \frac{v^2}{f^2} \frac{s^2 (c^2-s^2)-s^4}{2}\right)~, 
\end{equation} 
and for the neutral current we obtain 
\begin{equation} 
-Z_\mu \sqrt{g^2 + g'^2} \left( J^{0\mu} (1 + \frac{v^2}{f^2} 
\frac{c^2 s^2 - s^4 + c'^2 s'^2 - s'^4}{2}) + J^\mu (\frac{g'^2}{g^2 + g'^2} + 
\frac{v^2}{f^2} \frac{s'^2 (c'^2-s'^2)}{2})\right)~. 
\end{equation} 
In this case the masses of $Z$- and $W$-bosons are given by 
\begin{equation} 
m_Z^2 = (g^2 + g'^2) \frac{v^2}{4} \left(1 + 4 \frac{v_1'^2}{v^2}\right) 
\label{massz}
\end{equation} 
\begin{equation} 
m_W^2 = \frac{g^2 v^2}{4} \left(1 + 2 \frac{v_0'^2 + v_1'^2}{v^2}\right)~,
\label{massw} 
\end{equation} 
where $v_0'$ stands for a vev for the triplet with hypercharge $Y = 0$
and $v_1'$ for the triplet with hypercharge $Y=1$.
\FIGURE[ht!]{ 
\epsfig{file=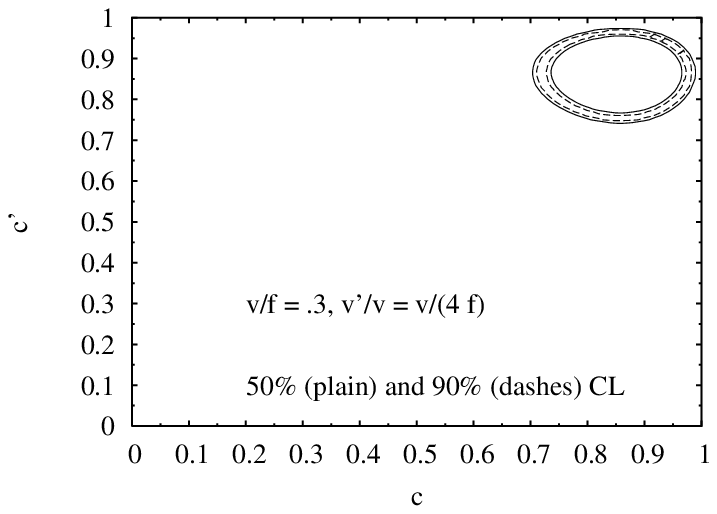,width=0.45\textwidth}
\epsfig{file=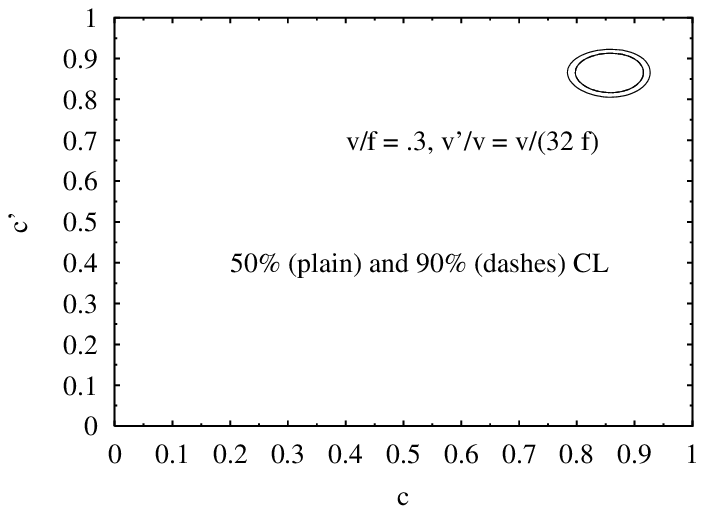,width=0.45\textwidth}\\
\epsfig{file=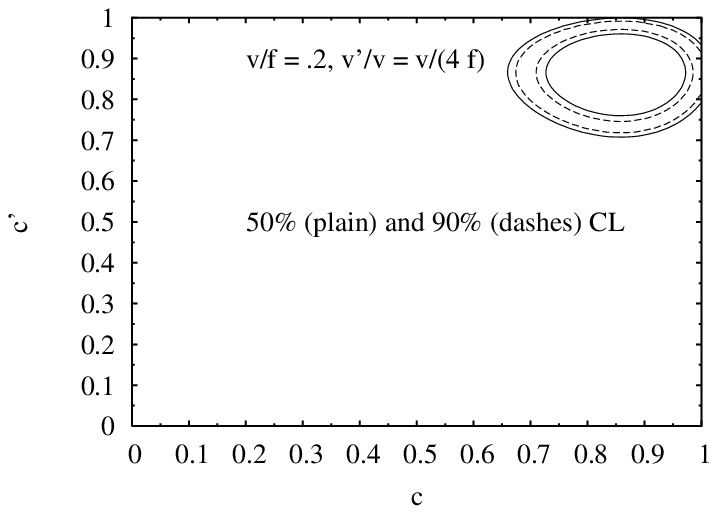,width=0.45\textwidth}
\epsfig{file=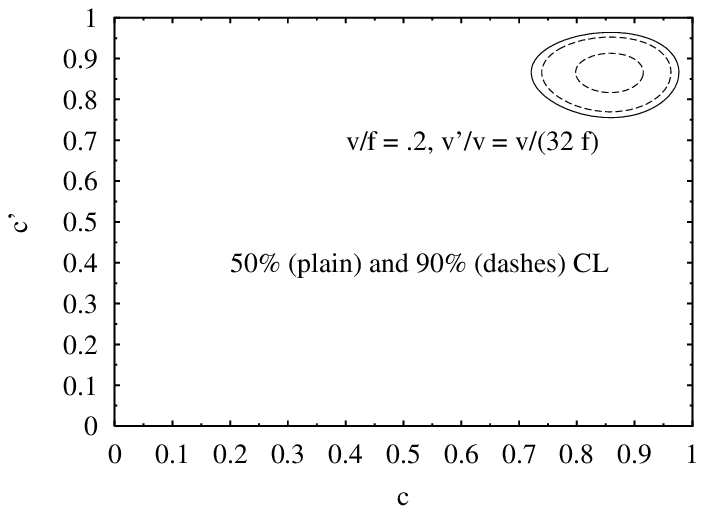,width=0.45\textwidth}\\
\epsfig{file=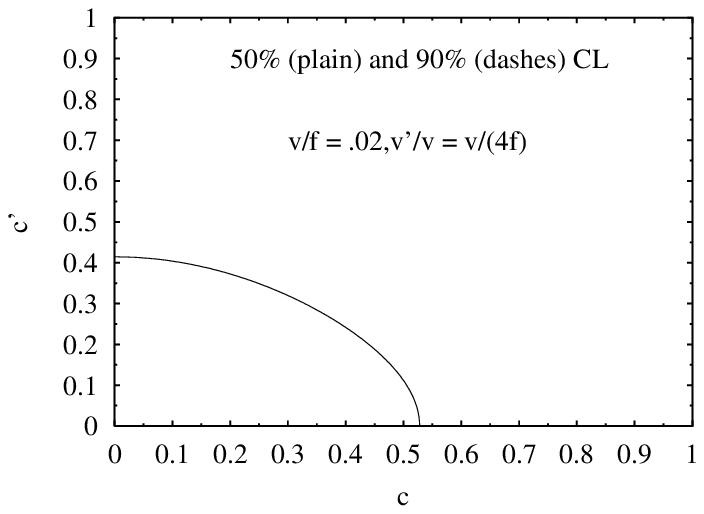,width=0.45\textwidth}
\epsfig{file=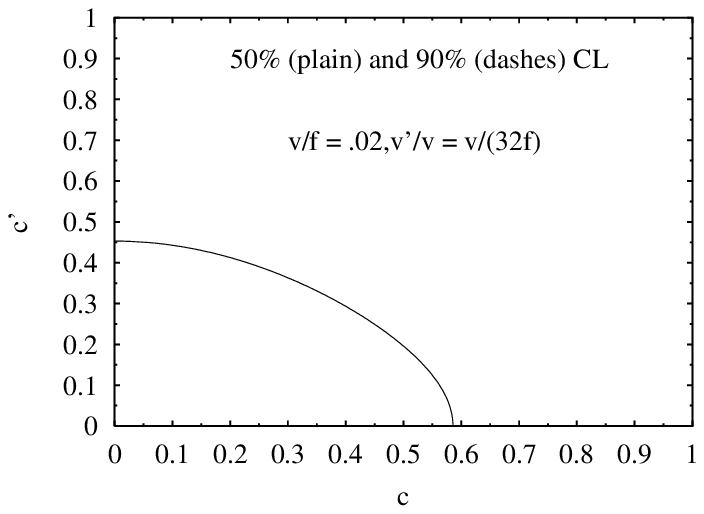,width=0.45\textwidth}
\caption{\label{epscustfig}\it 90\% and 50\% CL exclusion contours in the plane 
$c$-$c'$ of the cosines of the two mixing angles for three values of 
the ratio $v/f$ of the vev's of the $SO(9)/[SO(5)\times SO(4)]$ 
model. The allowed region lies  
inside the 90\% and 50\% bands, respectively.} 
} 
Note that, in contrast to the littlest Higgs model discussed in
Sec.~\ref{epsilons}, the masses of $Z$- and $W$-bosons are only
modified with respect to their standard (tree level) model by the
triplet vevs.  If the triplet vevs are similar in magnitude, custodial
symmetry violating effects on the $\rho$ paramter remain small at all
orders. This is a consequence of the approximate custodial symmetry
of the model.
 
The corrections to the $\epsilon$ 
parameters to the order $v^2/f^2$ are 
\begin{eqnarray} 
\epsilon_1 &=& \frac{v^2}{4f^2} \left[ 4 s^{\prime\, 2} 
\left( c^{\prime\, 2}-s^{\prime\, 2}\right) +2 c^2 s^2 -s^4 \right] 
+ 2 \frac{v'^2}{v^2}\\ 
\epsilon_2 &=& \frac{v^2}{4 c_{2\theta}\, f^2} \Big[ 4 s^{\prime\, 
2} \left( c^{\prime\, 2}-s^{\prime\, 2}\right) c_{\theta}^2 
c_{2\theta} +2 s^2 \left( c^2 -s^2 \right)\left( c^4_\theta -3 
c^2_\theta s^2_\theta +2 c^2_\theta -s^2_\theta \right) 
\nonumber \\ 
&+&s^4 (c^4_\theta + s^4_\theta) \Big]\\ 
\epsilon_3 &=& \frac{v^2}{2 s^2_{\theta}\, f^2} 
\left[ s^2 \left( c^2 -s^2 \right) 
\left( -c_{2\theta} + 2 s_\theta^2 c_\theta^2 \right) 
-s^4 c^2_\theta s^2_\theta \right]  
-\frac{2 c^2_{\theta}}{s^2_{\theta}} \frac{v'^2}{v^2 }~, 
\label{epsilonscustodial} 
\end{eqnarray} 
where we have again used the definition of $s_\theta$ and $c_\theta$
via Eq.~\ref{weinberg}. The effective triplet vev, $v'$, has been
defined as $v' = \sqrt{v_0'^2-v_1'^2}$.

The results of the analysis are illustrated in Fig.~\ref{epscustfig}
for three different values of $v/f$. The left panels correspond to
$v'/v = v/(4 f)$ and the right panels to $v'/v =v/(32
f) $, respectively.  Note that in principle $v'^2/v^2$ could also
become negative (if $v'_0 < v'_1$), but this possibility is almost
excluded by the data as in particular the constraint on $\epsilon_3$
becomes difficult to satisfy. The allowed region lies inside the
bands.  Similarly to the littlest Higgs model the allowed region
increases first with decreasing $v/f$ and disappears completely upon
reaching some limiting value.  The latter is almost the same as in the
littlest Higgs model and corresponds to the SM limit. However, in
contrast to the littlest Higgs model we find reasonable agreement
already for rather large values of $v/f$ for not too large values of
$v'/v$.  This clearly shows the enhanced custodial symmetry of the
model which makes it easier to satisfy the experimental constraint on
$\epsilon_1$.  One can argue that the two different triplet vevs
should be similar in size and at least partially compensate their
effects and that consequently $v'/v$ always remains small. A precise
evaluation of this effect is, however, not possible in the effective
theory since there are unknown order one factors in the radiatively
generated potential.  As can be already inferred from the expression
of the $W$- and $Z$-masses, Eqs.~(\ref{massz},\ref{massw}), a large
value for the difference of the triplet vevs spoils the custodial
symmetry. Therefore we find qualitative changes in the results when
varying $v'/v$. If the difference becomes too large, the constraints
on $\epsilon_1$ can no longer be satisfied easily and much more fine
tuning is needed in order to remain consistent with existing
experimental data.
 
\section{Low energy precision data} 
\label{lowenergy}  
Precision experiments at low energy allow a 
precise determination of the $g-2$ of the muon and of the weak charge 
of cesium atoms. We will analyse these data in order to see whether 
they can put constraints on our models. In a first step we will 
examine $g-2$ within the littlest Higgs model. To that end we will use 
a somewhat different technique than in the previous sections. Instead 
of deriving an effective Lagrangian by integrating out the heavy 
degrees of freedom we will use the linearized version of the model as 
presented in Ref.~\cite{Han:2003wu} and explicitly include 
corrections from the heavy bosons.  
\subsection{$g-2$ of the muon in the littlest Higgs model} 
We can use the results of Ref.~\cite{Leveille78} to calculate the 
corrections to $g-2$ of the muon which we will denote by $a_\mu$. The 
relevant contributions are discussed in App.~\ref{details}. 
 
\begin{figure}[ht] 
\begin{center} 
\epsfig{file=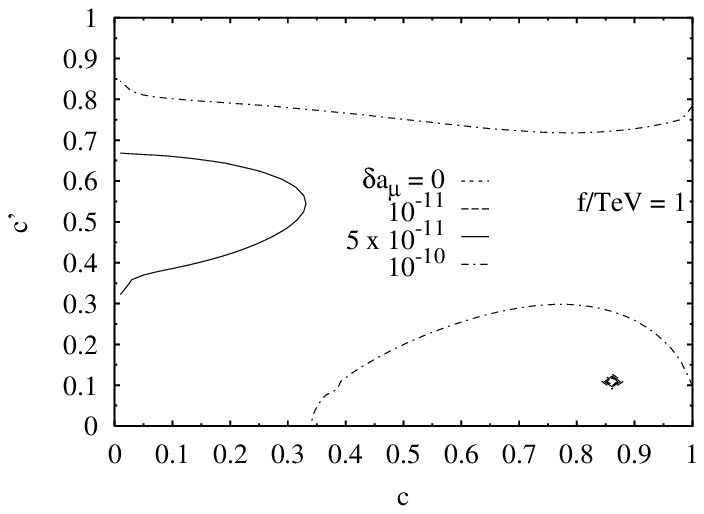,width=0.45\textwidth} 
\epsfig{file=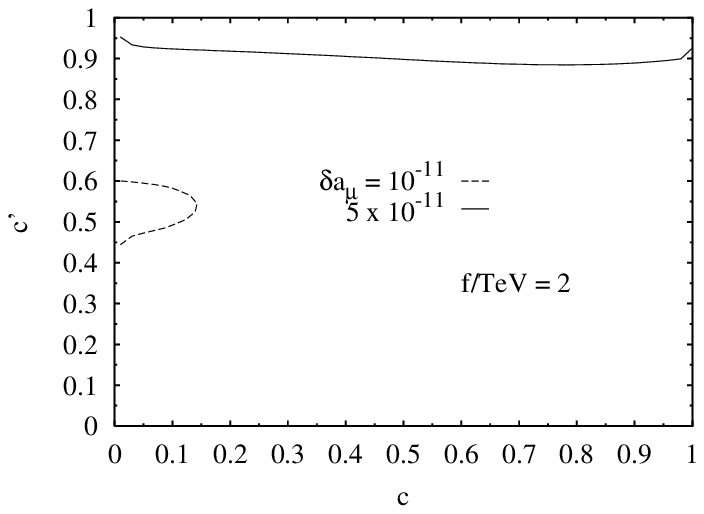,width=0.45\textwidth}\\ 
\caption{\label{fig2}\it Corrections to $g-2$ of the muon as a 
function of $c,c'$ for $v'^2/v^2 = v^2/(24f^2)$.} 
\end{center} 
\end{figure}

The difference between experiment and the standard model 
prediction for $a_\mu$ is~\cite{Groote:2003kg} 
\begin{equation} 
\delta a_\mu = a_\mu^\mathit{exp} - a_\mu^\mathrm{SM} = 17(18) \times 
10^{(-10)}~. 
\end{equation} 
The numerical results within the littlest Higgs model are relatively 
insensitive to the choice of parameter values of the model. We obtain 
a difference from the standard model value of at most $\delta a_\mu = 
a_\mu^\mathrm{LH} - a_\mu^\mathrm{SM}$ of the order of $1\times 
10^{-10}$. The contributions of the additional heavy particles are 
thereby completely negligible and the dominant contributions arise 
from the corrections to the light $Z$ and $W$ couplings. Thus the 
analysis of Ref.~\cite{Park:2003sq} is not complete. In 
Fig.~\ref{fig2} we display $\delta a_\mu$ for two different values of 
the symmetry breaking scale $f$ as a function of the cosines of the 
mixing angles $c,c'$.  For larger values of $f$ the corrections become 
even smaller. We thereby took the Higgs mass to be 113 GeV and used as 
experimental input $m_Z, \alpha$, and $G_F$ as before. A variation of 
the triplet vev does not change these findings.  The results in the 
model with custodial symmetry are in general closer to the SM limit 
than in the littlest Higgs model. Thus we expect even smaller 
corrections in that case and we shall not give an explicit evaluation 
of $g - 2$ within the model with approximate custodial symmetry. 
 
\subsection{Weak charge of cesium atoms} 
At low energy, parity violation in atoms is due to the electron-quark 
effective Lagrangian 
\begin{equation} 
\mathcal{L}_\mathit{eff} = \frac{G_F}{\sqrt{2}} (\bar{e} \gamma_\mu 
\gamma_5 e) (C_{1u} \bar{u} \gamma^\mu u + C_{1d} \bar{d} \gamma^\mu 
d)~. 
\label{Leff} 
\end{equation} 
The experimentally measured quantity is the so-called ``weak charge'' 
defined as 
\begin{equation} 
Q_W = -2 \left( C_{1u} (2 Z + N) + C_{1d} (Z + 2 N)\right)~, 
\end{equation} 
where Z, N are the number of protons and neutrons of the atom, 
respectively. 
 
\begin{figure}[ht] 
\begin{center} 
\epsfig{file=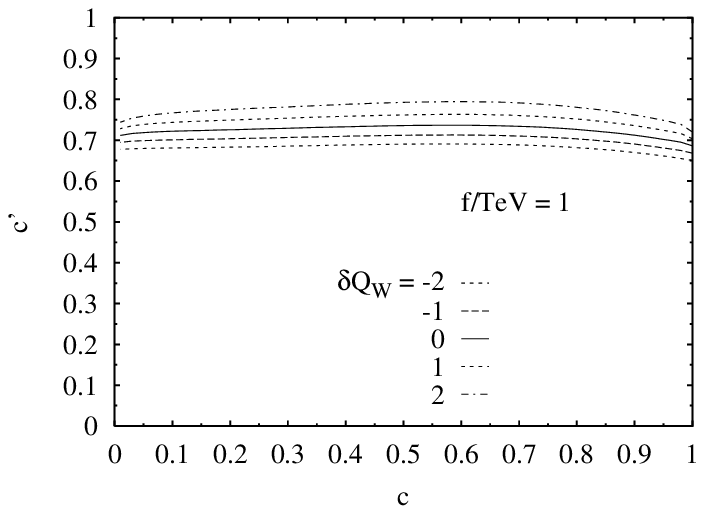,width=0.45\textwidth} 
\epsfig{file=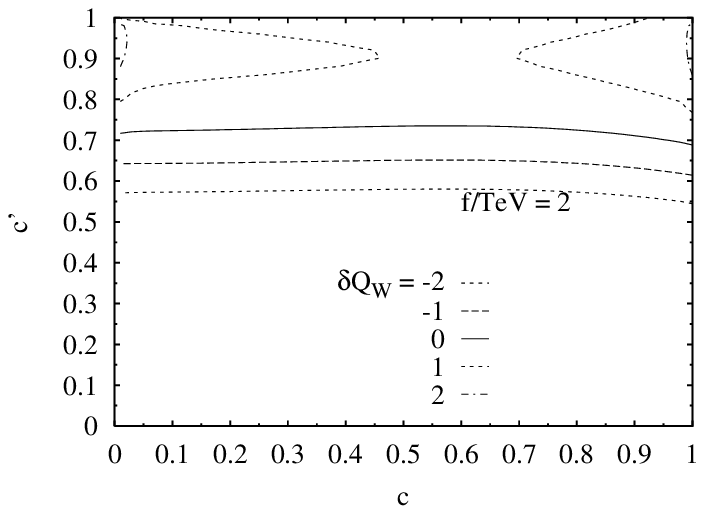,width=0.45\textwidth}\\ 
\epsfig{file=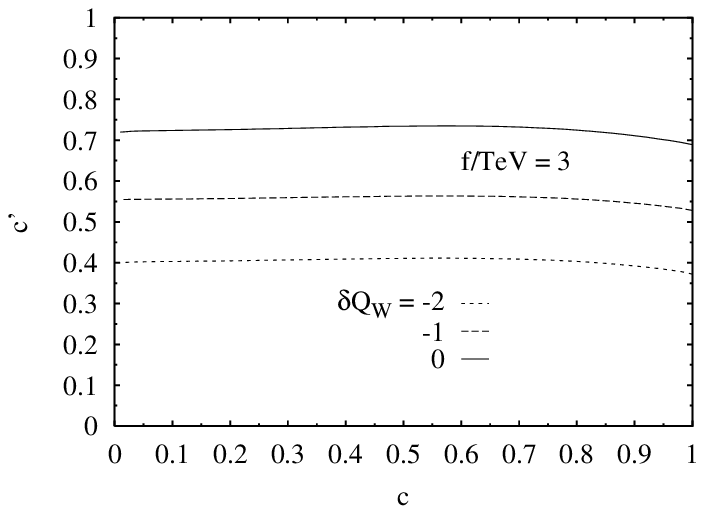,width=0.45\textwidth} 
\epsfig{file=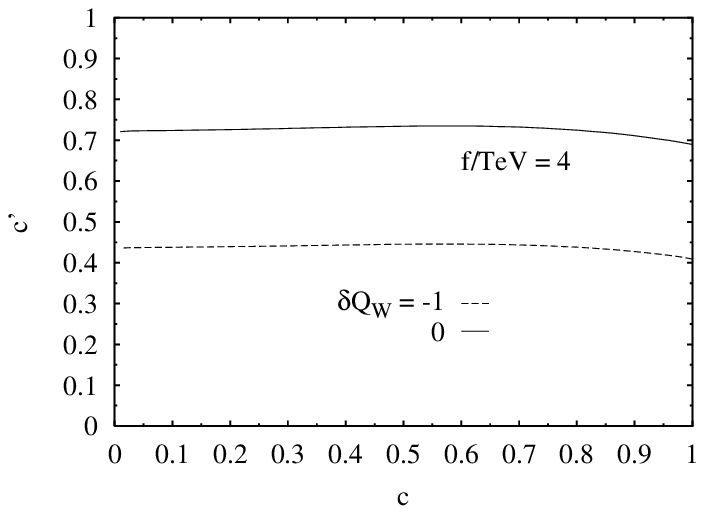,width=0.45\textwidth} 
\caption{\label{fig3}\it Corrections to the weak charge of cesium 
atoms as a function of $c$ and $c'$ in the 
littlest Higgs model.} 
\end{center} 
\end{figure} 
The effective Lagrangian, Eq.~\ref{Leff}, can be derived from the 
interaction of $Z, Z_H,$ and $A_H$ with the fermions by integrating 
out the heavy degrees of freedom.  The corresponding expressions for 
the littlest Higgs model as well as for the model with approximate 
custodial symmetry are given in Appendix~\ref{appb}. 
 
Recently precise data on cesium atoms have been reported in 
Ref.~\cite{Rosner:2001ck}: 
\begin{equation} 
Q_W(Cs)^\mathit{exp} = -72.2 \pm 0.8~. 
\end{equation} 
The standard model prediction is~\cite{Marciano:1990dp} 
\begin{equation} 
 Q_W(Cs)^\mathrm{SM} = -73.19 \pm 0.13~. 
\end{equation} 
Thus 
\begin{equation} 
\delta Q_W(Cs) = Q_W(Cs)^\mathit{exp} - Q_W^\mathrm{SM} = 0.99 \pm 
0.93~. 
\end{equation} 
The difference of the weak charge of Cs in the littlest Higgs model 
and the standard model is shown in Fig.~\ref{fig3} for different 
values of $f$ in the littlest Higgs model and in Fig.~\ref{fig3cust} 
for the model with approximate custodial symmetry. As experimental 
input for our analysis we have again used $m_Z, G_F,$ and $\alpha$. 
In order to discuss the weak charge result, let's consider the value  
$\delta Q_W(Cs)=1$ which is close to the present experimental central  
value. It is clear from Fig.~\ref{fig3} and \ref{fig3cust} that the value  
of the high scale $f$ should be in the range of few TeV in order to obtain  
the measured deviation. The allowed scale is slightly lower in the  
custodial model with respect to the non-custodial one as the custodial model  
is closer to the standard model in its predictions. When the scale $f$ is too  
large the new physics effects become negligible.  
The scale $f$ in the few TeV range is consistent with what is expected on the  
model-building side and from the LEP data for little Higgs models.  
Obviously this result should be taken only as a first indication as the error 
on $\delta Q_W(Cs)$ is large. 
 
\FIGURE[ht!]{ 
\epsfig{file=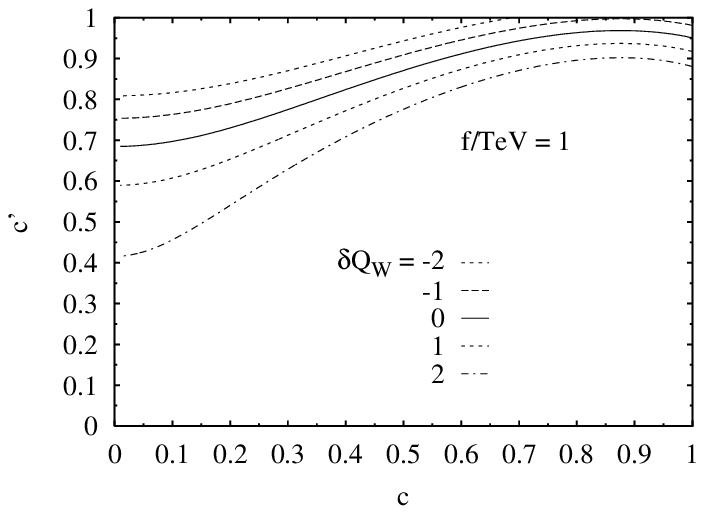,width=0.45\textwidth} 
\epsfig{file=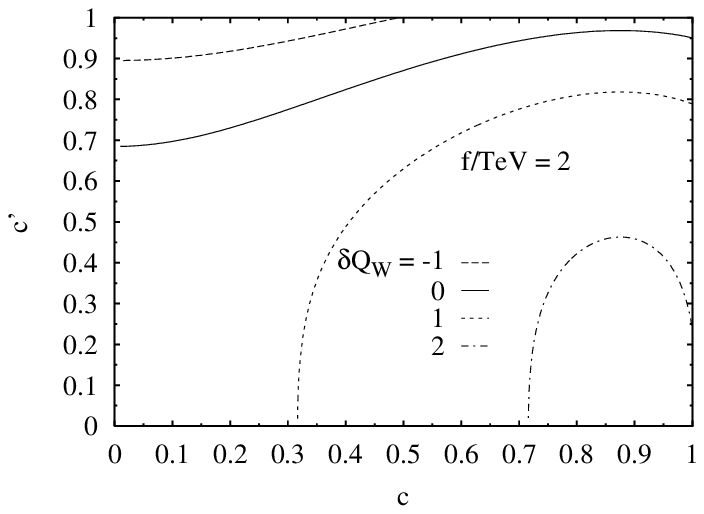,width=0.45\textwidth}\\ 
\epsfig{file=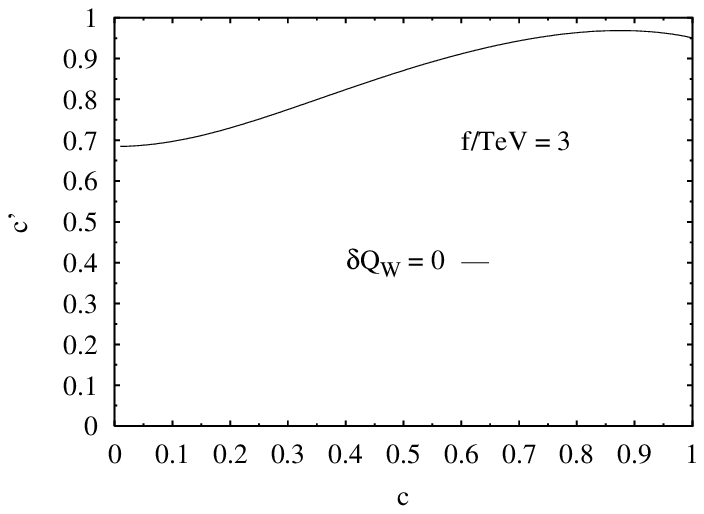,width=0.45\textwidth} 
\caption{\label{fig3cust}\it Corrections to the weak charge of cesium 
atoms as a function of $c$ and $c'$ in the 
little Higgs model with approximate custodial symmetry.} 
} 
\section{Conclusions} 
 
In this paper we have studied the low energy limit of a Little Higgs
model incorporating an approximate custodial symmetry. For
illustrational purposes we briefly presented in Sec.~\ref{epsilons}
the corresponding analysis for the ``littlest Higgs'' model which has
no custodial symmetry. In order to study the constraints coming from
the LEP/SLC experiments we have used a method which consists in
eliminating the heavy degrees of freedom.  We find, in agreement with
earlier studies in the literature~\cite{othlh,Han:2003wu}, rather
stringent limits on the littlest Higgs model imposed by existing
electroweak precision data.  This is mainly due to the difficulty of
the model to accommodate for the experimental results of the $\rho$
parameter.  Our main focus lied on the study of a model where
custodial symmetry is approximately fulfilled. As long as the value of
the effective Higgs triplet vev, which violates custodial symmetry,
does not become too large, we need much less fine tuning than in the
littlest Higgs model in order to satisfy the experimental
constraints. As the effective triplet vev is related to the difference
of the vevs for the $Y=1$ and the $Y=0$ Higgs triplet, this can always
be achieved if the two vevs are of similar size. Custodial
symmetry seems to be an essential ingredient for little Higgs models.
 
In the second part of this paper we look at the constraints from low 
energy precision data., i.e., $g-2$ of the muon and to the atomic 
"weak charge" of the cesium. To that end we apply a slightly different 
method: To evaluate the corrections to these quantities the 
contributions of the heavy degrees of freedom have directly been taken 
into account. 
The analysis of the low energy precision data does 
not change the above conclusions. For $g-2$ of the muon the 
corrections are simply too small to impose any new constraints on the 
model parameters. The actual state of precision for the weak charge 
does not allow for establishing new constraints either, even if the 
corrections are not negligible. 
 
\acknowledgments 
We thank S. Chang for valuable discussions on the
custodial little Higgs model.  M.O. wishes to thank the theory group
of IPN Lyon for the kind hospitality during the preparation of this
work.
 
\begin{appendix} 
\section{Details for the calculation of $g-2$ of the muon in the Littlest Higgs model} 
\label{details} 
\FIGURE[ht]{ 
\hspace{3cm}\epsfig{file=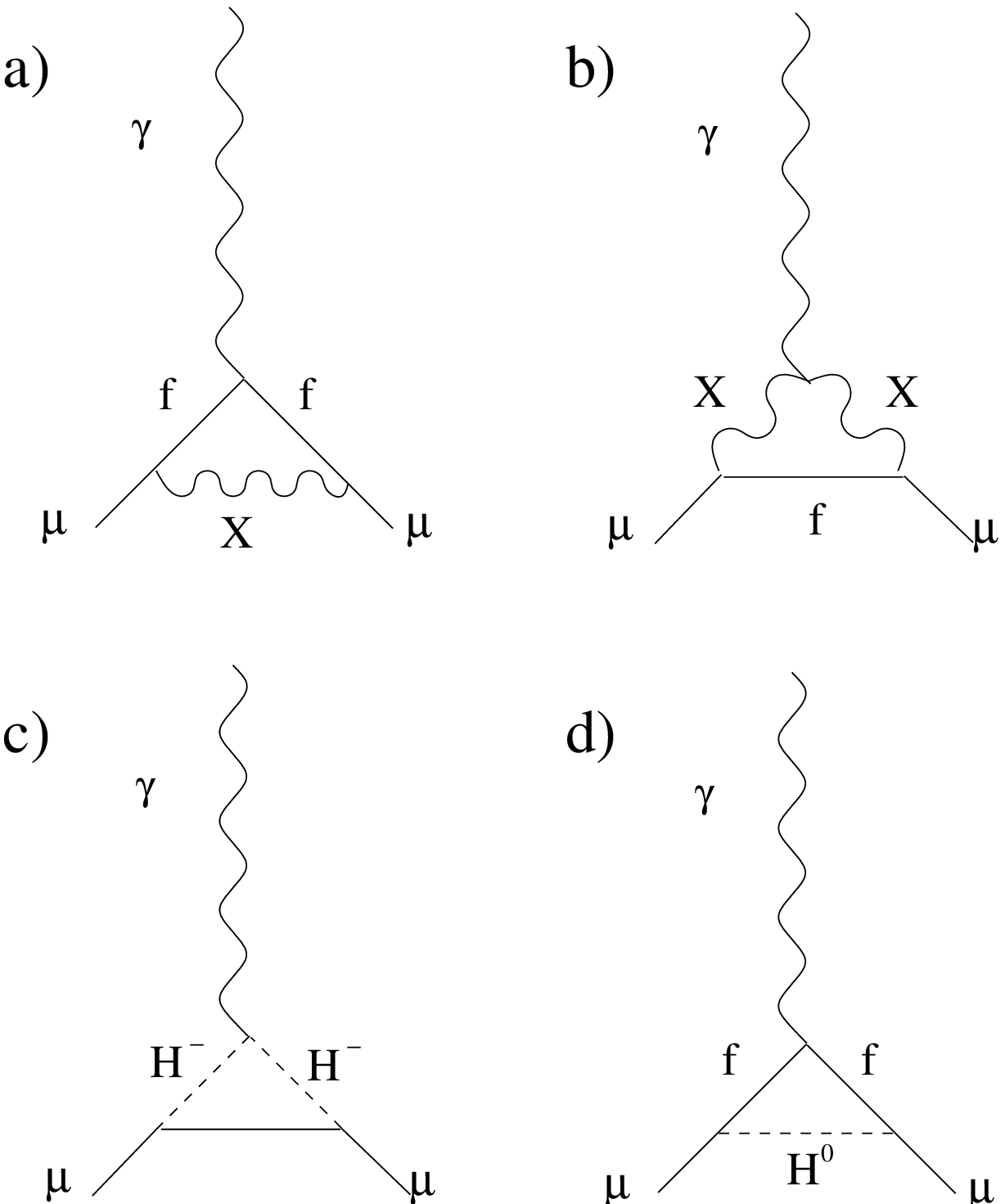,width=8cm}\hfill 
\caption{\label{fig:gm2}\it 
Loop graphs contributing to the weak correction to $\Delta g$. a) and b) 
correspond to the exchange of a vector boson $X$ while c) and d) are the 
Higgs sector contributions.} 
} 
The relevant one-loop Feynman 
diagrams are shown in Fig.~\ref{fig:gm2}. 
For graph a we have contributions from the exchange of a light and 
a heavy $Z$ and a light and a heavy photon. Since we have no 
flavour mixing interaction the fermion in the intermediate state 
can only be a muon. The contribution of the photon is not modified 
with respect to its standard model value (note that 
$U(1)_\mathit{em}$ is not broken so that this should be the 
case). 
 
Explicitly we have 
\begin{alignat}{1} 
[a_\mu]_a =& \frac{1}{8 \pi^2} \Bigg\{ C_V^2 
\left[ \frac{z}{3} + z^2\ \left( \frac{25}{12} + \ln(z)\right) + z^3\ \left( 
\frac{97}{10} + 6\ \ln(z)\right) + 
    z^4\ \left( \frac{208}{5} + 28\ \ln(z)\right) \right] \nonumber \\  
+& C_A^2 \left[ -5\ \frac{z}{3}  - z^2\ \left( \frac{19}{12} + \ln(z)\right)  
- z^3\ \left( \frac{77}{15} + 4\ \ln(z)\right) - z^4\ \left( \frac{173}{10}  
+ 14\ \ln(z)\right) \right]\Bigg\}\nonumber \\   
+& \mathcal{O}(z^5)~, 
\end{alignat} 
where $z = m_\mu^2/M^2_g$. With $M_g$ we denote the mass of the 
exchanged gauge boson, i.e., $m_{Z}, M_{Z_H} $ or 
$M_{A_H}$. $C_V$ and $C_A$ can be extracted from the vector and 
axial couplings of the corresponding gauge bosons to the muons, 
see Ref.~\cite{Leveille78}. The corresponding expressions are 
listed in appendix~\ref{appa}. 
 
For graph b we obtain contributions from light and heavy charged $W$ 
bosons. The intermediate fermion has to be neutral, i.e., it is a 
neutrino. If we neglect the mass of the neutrino we obtain 
\begin{equation} 
[a_\mu]_b =\frac{C_V^2}{4 \pi^2}  \left(5 \frac{z}{3} + \frac{z^2}{3} + 3 
\frac{z^3}{20} + \frac{z^4}{12}\right) + \mathcal{O}(z^5)~, 
\end{equation} 
where we note again $z = m_\mu^2/M_g^2$. $M_g$ can be the mass of the 
light or the heavy $W$ bosons. We also used the fact that $C_V = - C_A$. 
 
Graph c receives contributions from singly charged scalars and a 
neutrino in the intermediate states. There are no contributions from 
doubly charged scalars since they do not couple to the corresponding 
fermions. Neglecting again the neutrino mass the result can be written 
\begin{equation} 
[a_\mu]_c = \frac{C_S^2}{8 \pi^2} \left(\frac{z}{6} + \frac{z^2}{12} + 
\frac{z^3}{20} + \frac{z^4}{30} \right) + \mathcal{O}(z^5)~, 
\end{equation} 
with $z = m_\mu^2/M_H^2$. The pseudoscalar coupling ($C_P$) vanishes. 
 
For graph d we have to consider muons and a neutral scalar as 
intermediate states. The result is 
\begin{alignat}{1} 
[a_\mu]_d &=\frac{1}{8 \pi^2} \Bigg\{ -C_S^2 
\Bigg[ z \ \left( \frac{7}{6} + \ln(z)\right) + z^2\ \left( \frac{39}{12}  
+ 3 \ \ln(z)\right) + z^3\ \left( \frac{201}{20} + 9\ \ln(z)\right)  
\nonumber \\ 
&+ z^4\ \left( \frac{484}{15} + 28\ \ln(z)\right) \Bigg] 
+ C_P^2 \Bigg[ z\ \left( \frac{11}{6} + \ln(z)\right) + z^2\ \left(  
\frac{89}{12} + 5\ \ln(z)\right) \nonumber \\  
&+ z^3\ \left( \frac{589}{20} 
+ 12\ \ln(z)\right) + z^4\ \left( \frac{1732}{15} + 84\ 
\ln(z)\right) \Bigg] \Bigg\} + \mathcal{O}(z^5)~, 
\end{alignat} 
where $z = m_\mu^2/M_h^2$, with $M_h$ being the mass of the neutral scalar. 
\section{Couplings for the calculation of $g-2$ in the littlest Higgs model} 
\label{appa} 
The vector and axial vector couplings are given as follows. For the 
$Z$ we obtain up to the order $\mathcal{O}({v^2}/{f^2})$: 
\begin{eqnarray} 
C_V &=& \frac{g}{2 c_\theta} \left[ 2 s_\theta^2 - \frac{1}{2} +  
\frac{v^2}{f^2} \left( 3 s_\theta \frac{x_Z^B}{s' c'} \left(  
\frac{c'^2}{2} -\frac{1}{5}\right) - c_\theta x_Z^W \frac{c}{2 s}\right) 
\right]\\ 
C_A &=& \frac{g}{2 c_\theta} \left[ \frac{1}{2} + \frac{v^2}{f^2} \left(  
s_\theta \frac{x_Z^B}{s' c'} \left( \frac{c'^2}{2} -\frac{1}{5}\right)  
+ c_\theta x_Z^W \frac{c}{2 s}\right)\right]~. 
\end{eqnarray} 
For the heavy $Z$ we obtain: 
\begin{eqnarray} 
C_V &=& C_A = \frac{g c}{4 s}~. 
\end{eqnarray} 
Note that it is sufficient to retain the leading order contributions 
for the couplings of the heavy bosons since their contributions to 
$g-2$ are already suppressed by one order in $v^2/f^2$ due to 
the mass. The heavy photon couplings are given by 
\begin{eqnarray} 
C_V &=& 3 \frac{g'}{2 s' c'} \left( \frac{c'^2}{2} - \frac{1}{5}\right)\\ 
C_A &=&  \frac{g'}{2 s' c'} \left(\frac{c'^2}{2} - \frac{1}{5}\right)~. 
\end{eqnarray} 
For the $W$ bosons we obtain: 
\begin{eqnarray} 
C_V &=&  -C_A = \frac{g}{2 \sqrt{2}} \left(1-\frac{v^2}{f^2} \frac{c^2}{2} (c^2-s^2)\right)~, 
\end{eqnarray} 
and for the heavy $W$-bosons: 
\begin{eqnarray} 
C_V &=& -C_A = \frac{g c}{2 \sqrt{2} s}~. 
\end{eqnarray} 
The relevant Higgs couplings are given by 
\begin{eqnarray} 
C_S &=& \frac{m_\mu}{\sqrt{2} v} \left(\frac{v}{f}-4 \frac{v'}{v}\right)\\ 
C_P &=& 0 
\end{eqnarray} 
for the heavy Higgs and 
\begin{eqnarray} 
C_S &=& \frac{m_\mu}{v} \left( 1-4 \frac{v'^2}{v^2}- \frac{2}{3}  
\frac{v^2}{f^2}\right)\\ 
C_P &=& 0~. 
\end{eqnarray} 
for the light higgs which contributes only to graph d) (see 
Fig.~\ref{fig3}). Note that there are no leading order contributions 
to the couplings of the heavy Higgs, i.e., at the order we consider 
the heavy Higgs contribution vanishes. 
 
\section{Parameters of the effective Lagrangian for the weak charge} 
\label{appb} 
In the littlest Higgs model we obtain 
\begin{eqnarray} 
C_{1u}^{A_H} &=& -\frac{\sqrt{2}}{M_{A_H}^2 G_F}  
\frac{\alpha \pi}{ c_\theta^2 s'^2 
c'^2} \left(-\frac{1}{5} + \frac{c'^2}{2}\right) 
\left(\frac{1}{3}-\frac{5}{6} c'^2\right)  \\ 
C_{1d}^{A_H} &=& -\frac{\sqrt{2}}{M_{A_H}^2 G_F}  
\frac{\alpha \pi}{c_\theta^2 s'^2 
c'^2} \left(-\frac{1}{5} + \frac{c'^2}{2}\right) 
\left(-\frac{1}{15}+\frac{1}{6} c'^2\right) \\ 
C_{1u}^{Z_H} &=&-C_{1d}^{Z_H} =  -\frac{\sqrt{2}}{M_{Z_H}^2 G_F}  
\frac{\alpha \pi c^2}{4 s_\theta^2 s^2}\\ 
C_{1u}^{Z} &=& -\frac{\sqrt{2}}{m_{Z}^2 G_F}  
\frac{\alpha \pi}{s_\theta^2 c_\theta^2} 
\Bigg\{\frac{1}{4} -\frac{2}{3} s_\theta^2 + \frac{v^2}{f^2}\Bigg[ \left( 
\frac{1}{2} -\frac{4}{3} s_\theta^2\right) \left( c_\theta x_Z^{W'}  
\frac{c}{2s} + \frac{s_\theta x_Z^{B'}}{s' c'} 
\left(-\frac{1}{5} + \frac{c'^2}{2}\right)\right) \nonumber \\  
&& + \frac{1}{2} \left(c_\theta x_Z^{W'} \frac{c}{2s} +  
\frac{s_\theta x_Z^{B'}}{s' c'} 
\left( -\frac{1}{3} -\frac{c'^2}{6}\right)\right) \Bigg] \Bigg\}\\ 
C_{1d}^{Z} &=& -\frac{\sqrt{2}}{m_{Z}^2 G_F}  
\frac{\alpha \pi}{s_\theta^2 c_\theta^2} 
\Bigg\{ -\frac{1}{4} +\frac{1}{3} s_\theta^2 + \frac{v^2}{f^2}\Bigg[ \left( 
-\frac{1}{2} +\frac{2}{3} s_\theta^2\right) \left(c_\theta x_Z^{W'}  
\frac{c}{2s} + \frac{s_\theta x_Z^{B'}}{s' c'} 
\left( -\frac{1}{5} + \frac{c'^2}{2}\right)\right) \nonumber \\ &&  
+ \frac{1}{2} \left( -c_\theta x_Z^{W'} \frac{c}{2s} +  
\frac{s_\theta x_Z^{B'}}{s' c'} 
\left( -\frac{1}{15} + \frac{c'^2}{6}\right)\right) \Bigg] \Bigg\}~, 
\end{eqnarray} 
with 
\begin{eqnarray} 
x_Z^{W'} &=& -s\,c\,\frac{c^2-s^2}{2 c_\theta}\\ 
x_Z^{B'} &=& -5\,c'\,s'\,\frac{c'^2-s'^2}{2 s_\theta}~. 
\end{eqnarray} 
In the model with approximate custodial symmetry we get: 
\begin{eqnarray} 
C_{1u}^{A_H} &=& \frac{\sqrt{2}}{M_{A_H}^2 G_F} \frac{\alpha \pi}{3 c_\theta^2 
c'^2} \left(1 + \frac{s'^2}{4}\right) \\ 
C_{1d}^{A_H} &=& -\frac{\sqrt{2}}{M_{A_H}^2 G_F} \frac{\alpha \pi}{6 c_\theta^2 
c'^2} \left(1 - \frac{s'^2}{2}\right) \\ 
C_{1u}^{Z_H} &=& -C_{1d}^{Z_H} = -\frac{\sqrt{2}}{M_{Z_H}^2 G_F}  
\frac{\alpha \pi s^2}{4 s_\theta^2 c^2} \\ 
C_{1u}^{Z} &=& -\frac{\sqrt{2}}{m_{Z}^2 G_F} \frac{\alpha\pi} 
{s_\theta^2 c_\theta^2} 
\Bigg[ \frac{1}{4} -\frac{2}{3} s_\theta^2 \\ &+& 
\frac{v^2}{24 f^2}(-2 (c'^2-s'^2) + s'^2 (c'^2-s'^2) (1-4 s_\theta^2) 
- s^2 (c^2-s^2) (1-4 c_\theta^2))\Bigg] \nonumber \\ 
C_{1d}^{Z} &=& -\frac{\sqrt{2}}{m_{Z}^2 G_F} \frac{\alpha\pi} 
{s_\theta^2 c_\theta^2} 
\Bigg[ \frac{1}{3} s_\theta^2 -\frac{1}{4}+ \frac{v^2}{24 f^2}(-s^2 (c^2-s^2) 
(1 + 2 c_\theta^2) \nonumber \\ 
&+& (c'^2-s'^2) (1- 2 c_\theta^2 s'^2) )\Bigg]~. 
\end{eqnarray} 
Note that the expressions do not depend on the triplet vevs. 
\end{appendix} 
 
 
\end{document}